\documentclass[twocolumn, 10pt, journal]{IEEEtran}
\usepackage{amsmath}
\usepackage[pdftex]{graphicx}
\usepackage{graphicx}
\usepackage{pstricks}
\usepackage{amsfonts}
\usepackage{tabularx}
\newtheorem{Lemma}{Lemma}
\newtheorem{Remark}{Remark}

\newtheorem{Proposition}{Proposition}

\usepackage{amssymb}
\usepackage{multirow}
\usepackage{epsfig}
\usepackage{epstopdf}

\usepackage[noadjust]{cite} 
\usepackage{algorithm,algpseudocode,graphicx}

\begin{document}
\title{Secure Transmission in NOMA-enabled Industrial IoT with Resource-Constrained Untrusted Devices}
\author{Sapna Thapar,~\IEEEmembership{Student Member,~IEEE,}  Deepak Mishra,~\IEEEmembership{Senior Member,~IEEE,} and 
\\ Ravikant Saini,~\IEEEmembership{Member,~IEEE} 

\thanks{S. Thapar and R. Saini are with the Department of Electrical Engineering, Indian Institute of Technology Jammu, Jammu, Jammu $\&$ Kashmir 181 221, India (e-mail: thaparsapna25@gmail.com; ravikant.saini@iitjammu.ac.in). 

D. Mishra is with the School of Electrical Engineering and Telecommunications, University of New South Wales, Sydney, NSW 2052, Australia (e-mail: d.mishra@unsw.edu.au).}

\thanks{This research work was supported by the Science and Engineering Research Board, DST, India under Grant CRG/2021/002464.}

}

\maketitle

\begin{abstract}
%
The security of confidential information associated with devices in the industrial Internet of Things (IIoT) network is a serious concern. This article focuses on achieving a non-orthogonal multiple access (NOMA)-enabled secure IIoT network in the presence of untrusted devices by jointly optimizing the resources, such as decoding order and power allocated to devices. Assuming that the devices are resource-constrained for performing perfect successive interference cancellation (SIC), we characterize the residual interference at receivers with the linear model. Firstly, considering all possible decoding orders in an untrusted scenario, we obtain secure decoding orders that are feasible to obtain a positive secrecy rate for each device. Then, under the secrecy fairness criterion, we formulate a joint optimization problem of maximizing the minimum secrecy rate among devices. Since the formulated problem is non-convex and combinatorial, we first obtain the optimal secure decoding order and then solve it for power allocation by analyzing Karush-Kuhn-Tucker points. Thus, we provide the closed-form global-optimal solution of the formulated optimization problem. Numerical results validate the analytical claims and demonstrate an interesting observation that the conventional decoding order and assigning more power allocation to the weak device, as presumed in many works on NOMA, is not an optimal strategy from the secrecy fairness viewpoint. Also, the average percentage gain of about $22.75\%$, $50.58\%$, $94.59\%$, and $98.16\%$, respectively, is achieved by jointly optimized solution over benchmarks ODEP (optimal decoding order, equal power allocation), ODFP (optimal decoding order, fixed power allocation), FDEP (fixed decoding order, equal power allocation), and FDFP (fixed decoding order, fixed power allocation).

\textit{Index Terms$-$}
Non-orthogonal multiple access, physical layer security, secrecy fairness, imperfect SIC, joint optimization.
\end{abstract}

\section{Introduction}\label{section1}
The adoption of the Internet of Things (IoT) in the industrial domain, i.e., the industrial IoT (IIoT), is rapidly becoming essential in creating hyper-connected cyber-physical networks in several verticals such as electricity, transportation, automation, healthcare, and manufacturing \cite{6714496}, \cite{9548837}. IIoT is a network of industrial devices connected to the Internet using state-of-the-art information and communications technologies to create a system that can capture, analyze, monitor, and exchange real-time data. However, due to the constraints of the scarce spectrum, connecting billions of IIoT devices in a wireless network is a challenge. Also, due to the broadcast nature of wireless transmission, the security of confidential information associated with IIoT devices is a severe concern \cite{8952830}, \cite{9169653}.Consequently, there has been an increase in research interest in secure data transmission from academia and industry in various scenarios, including multiple input multiple output (MIMO), \cite{9164916}, non-orthogonal multiple access (NOMA) \cite{nomasurvey}, intelligent reflecting surface \cite{9383283},  unmanned aerial vehicle assisted mobile edge computing networks \cite{9803857}, \cite{9954169}, etc.

As a favourable solution to realize massive connectivity over limited resources in IIoT, NOMA has recently drawn an increasing amount of research efforts \cite{nomasurvey}. It allows multiple devices to share the same resource block  (i.e., same time, frequency, and code), alleviating the spectrum shortage problem. For information confidentiality in wireless transmissions, physical layer security (PLS) techniques are recently emerging as a promising solution. The basic concept of PLS is to exploit the randomness of wireless channels and interference to enhance the signal reception at legitimate devices while reducing the signal reception at eavesdropping devices \cite{wyner1975wire}, \cite{PLS}. Thus, by incorporating NOMA and PLS, a spectrally-efficient and secure wireless communication network can be envisioned for IIoT.

\subsection{Related Works}\label{related_art}
Recently, many works have utilized NOMA to ensure massive connectivity requirements in different scenarios for IIoT \cite{8869734}, \cite{9674866}, \cite{8892479}. However, because of the broadcast nature of the wireless transmission channel, potential adversaries can cause a security risk to communication in NOMA-enabled IIoT networks. Many existing works have considered PLS techniques to protect NOMA networks in different scenarios. For example, in \cite{he2017design}, the authors optimized the resource allocation in terms of decoding order, power allocation, and data rate to protect the information associated with the intended device. In \cite{zhang2016secrecy}, the authors derived the optimal power allocation to maximize the achievable secrecy sum rate at devices for a single-input single-output NOMA network. In \cite{8309413}, a novel beamforming scheme that exploits artificial noise to improve the secrecy performance at devices in a multiple-input single-output network was proposed. In \cite{9763336}, a NOMA-assisted secure computation offloading was investigated under an eavesdropping attack, where a wireless user is paired with an edge-computing user to provide cooperative jamming to the eavesdropper while gaining the opportunity to transmit its data. In addition to eavesdropping, other attack modes like jamming were investigated in \cite{8187679}, where intelligent learning-based algorithms, such as Q-learning algorithms, were used to counter the intelligent attackers. Also, in \cite{7812773}, an artificial noise-aided beamforming approach was proposed to achieve secure communication in a large-scale NOMA network with randomly dispersed devices. 

The studies cited above \cite{he2017design}-\cite{7812773} were limited to the security issue of multiplexed NOMA devices against external eavesdropping devices. However, the multiplexed devices sharing the same resource block also can be potential eavesdroppers intercepting the confidential information of each other \cite{basepaper}, \cite{tvt_2020}. Therefore, we should consider an antagonistic network in which each device is assumed to be untrusted. An untrusted scenario is a hostile but realistic situation in which no device trusts others and wants to safeguard its own confidential information. As a result, it becomes essential to allocate resources in the network in such a way that the secrecy of each device is ensured against the other multiplexed untrusted devices, which is a relatively more complex problem.

Assuming the strong device (with better channel gain) as trusted and the weak device (with poorer channel gain) as untrusted for a two-device NOMA network, the authors derived the secrecy outage probability (SOP) for the strong device in \cite{basepaper}. Similarly, \cite{7833022} analyzed the sum secrecy rate of the strong devices against weak devices for a multiple-input single-output network. In \cite{9217169}, the SOP was investigated for the strong device against the untrusted weak device for a friendly jammer relay scenario. 
In contrast, \cite{9172088} considered the strong device as untrusted, and analyzed the optimal power allocation for a secure NOMA network by adjusting the order of successive interference cancellation (SIC) and utilizing a cooperative jammer. Likewise, in \cite{9004475}, a directional demodulation based method was proposed to secure the data of the weak device against the strong device. Furthermore, to safeguard the data of each NOMA device from the other, the authors proposed a linear precoding approach in \cite{9169675}. Similarly, to obtain a positive secrecy rate for each device against the other  device in a two-device NOMA-enabled network, an optimal decoding order was explored in \cite{tvt_2020}, and SOP and its optimization over power allocation were derived for each device. In \cite{pradosh_comm_letter}, the ergodic secrecy rate performance was analyzed for each possible decoding order in an untrusted NOMA network, and then, the optimal decoding order was identified. 

\subsection{Research Gap and Motivation}\label{gap_and_motivation}
Notwithstanding the gainful results in handling secrecy issues among untrusted devices in NOMA-enabled networks, several works, e.g. \cite{basepaper}, \cite{7833022}-\cite{9169675}, over-optimistically considered that the devices could perform perfect SIC. According to this ideal setup, the interference from previously decoded devices is fully subtracted when the signal associated with the later devices is decoded. Thus, the decoded devices do not interfere with other devices. This strong assumption makes the system model simple and might not be realistic. In a practical scenario, the devices are resource-constrained to perform perfect SIC due to various practical implementation issues in IIoT networks, such as hardware limitation, inaccurate calibration, estimation error, multiple types of noise, and complexity scaling \cite{7881111}. Consequently, imperfect SIC, where the residual interference (RI) from the formerly imperfectly decoded devices inevitably abides while decoding the signals of subsequent devices \cite{7881111}, should be taken at receivers while doing any investigation on NOMA. 

In the literature of NOMA, some research works considered the RI as a particular constant value \cite{constant_sic1}, \cite{constant_sic2},  referred to as the \emph{Constant RI Model}. In contrast, many other works took the RI as a linear function of the power assigned to the interfering signal \cite{7881111}, \cite{8755843}, referred to as the \emph{Linear RI Model}. Nevertheless, there seem to be fewer studies that have considered the impact of RI while handling the secrecy issue in untrusted NOMA networks. The authors in \cite{tvt_2020} analyzed the secrecy performance of devices in an untrusted NOMA network with imperfect SIC but considered the constant RI model. However, the RI obtained from decoded devices may not be a constant value in practice. The constant RI is a strong and unrealistic assumption that over-simplifies the model and leads to prediction errors. In contrast, realistic influence of imperfect SIC may be observed with the linear RI model since decoders' performance significantly depends on the interfering signal's power. \emph{Motivated by this solid observation, in this work, we mainly focus on obtaining a secure NOMA-enabled IIoT network in the presence of untrusted devices, considering the effect of imperfect SIC at receivers with a linear RI model.}

Note that \cite{pradosh_comm_letter} explored a secure NOMA network with a linear RI model, but the investigation was carried out for maximizing ergodic secrecy performance for each device. However, this article fills a significant gap in the literature by optimizing resources to maximize secrecy fairness among devices, which has not yet been studied in the literature. Note that fairness is an important performance metric in order to guarantee the achievable rates for weak devices, as considered in many works in the literature  \cite{tvt_2020}, \cite{9500064}. It is because focusing exclusively on the sum rate may result in substantial rate loss for weak devices, as the system tends to allocate most of the communication resources to strong devices when the sum rate is maximized. The weak devices may even be unable to be served in some extreme cases. Thus, in this work, the fundamental basis for studying secrecy fairness is that weak devices may also obtain enough communication resources similar to strong devices so that there is no loss in the achievable secrecy rate performance for weak devices. \emph{Therefore, by following \cite{tvt_2020}, \cite{9500064}, we in this work focus on maximizing secrecy fairness between devices, where we maximize the minimum secrecy rate between devices.}
 
To optimize the network's secrecy fairness performance, the decoding order and power allocation to the multiplexed devices are the two key parameters. In the literature, many research works are limited to the conventional decoding order of NOMA. However, we may change the decoding sequence for each device \cite{tvt_2020}, \cite{pradosh_comm_letter}, \cite{7343355}. The fact is that SIC is a physical layer capability that allows receiving ends to extract the
superimposed signal. Thus, any device can decode a signal of
itself or others at any stage resulting in various decoding orders. Besides, most existing literature assumes that NOMA is based on more power allocation to the device with weaker channel conditions, which is not true \cite{8823873}. Therefore, it would be interesting to investigate if the conventional approach of decoding order and power allocation is optimal from the secrecy fairness viewpoint. \emph{Encouraged by these substantial observations, in this work, we jointly optimize the resources, such as decoding order and power allocation, for maximizing the secrecy fairness between devices.}

\subsection{Main Contributions}\label{contributions}
The key contributions of this paper are summarized below: 
\begin{itemize}
\item We focus on achieving a secure NOMA-enabled IIoT network in the presence of untrusted devices, considering the real effect of imperfect SIC with a linear RI model. In this respect, we first find out the feasible power allocation condition to obtain a positive secrecy rate for each device in all possible decoding orders. This way, we identify the feasible secure decoding orders that can provide a positive secrecy rate for each device.
\item  We focus on optimizing the resources, such as secure decoding order and transmission power allocated to devices, from the perspective of secrecy fairness. Under the secrecy fairness criterion, we formulate and solve a joint optimization problem of maximizing the minimum secrecy rate between devices over a set of secure decoding orders and transmission power allocation. The formulated problem is combinatorial and non-convex. Therefore, we first find the optimal secure decoding order and then solve it over power allocation by obtaining candidates of optimal solution with Karush-Kuhn-Tucker (KKT) conditions. Thus, we provide the closed-form global-optimal solution to the formulated problem.
\item Lastly, we present numerical results to confirm the accuracy of the analysis, provide insightful discussion into the impact of network parameters on the optimal performance, and show the performance gains achieved by the optimal results over different benchmarks.
\end{itemize}

\textit{Notations}: Bold uppercase and lowercase letters, respectively, are used to refer to matrices and column vectors. We denote the $(u, v)$-th entries of matrix $\textbf{A}$ by $[\textbf{A}]_{u,v}$. The $u$-th entry of vector $\textbf{a}$ is indicated by $[\textbf{a}]_{u}$. 

\section{NOMA-enabled IIoT with Untrusted Devices}\label{section2}
In this section, firstly, we describe the network model. Then, we explain the fundamental principle of NOMA transmission. Further, we discuss all possible decoding orders for a NOMA-enabled network in the presence of untrusted devices. Lastly, we present the mathematical definition of the achievable secrecy rate for a device against the other untrusted device.

\subsection{Network Model}\label{network_model}
\begin{figure}[!t]
\centering
\includegraphics[scale=.36]{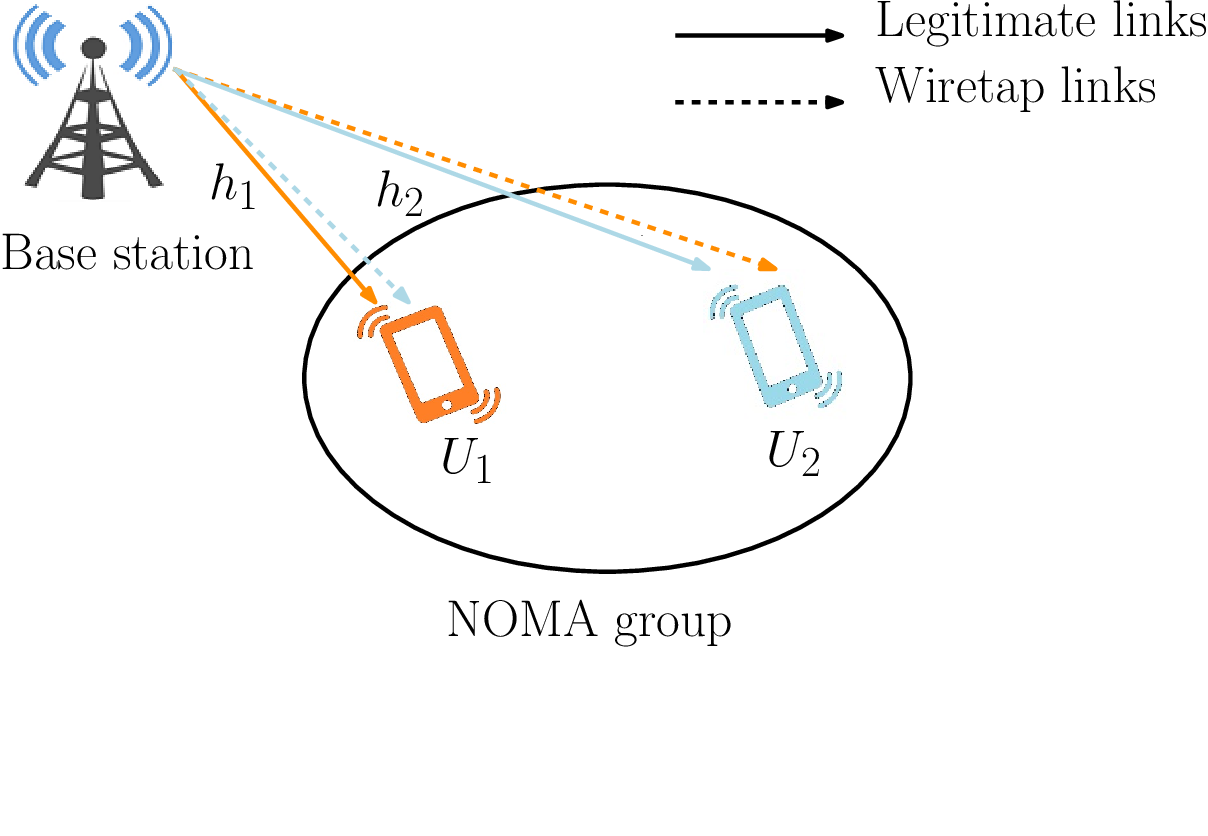}
\vspace{-0.8cm}
\caption{Illustration of a NOMA-enabled IIoT network with two untrusted devices. Each device may attempt to hear the information of the other device. }
\label{Figure1_networkmodel}
\vspace{-0.07cm}
\end{figure}

We consider a NOMA-enabled IIoT network, where the base station communicates with two devices, as depicted in Fig. \ref{Figure1_networkmodel}. In the network, both devices are assumed to be untrusted. The $n$-th device of the network is symbolized by $U_{n}$, where $n\in \mathbb{N}=\{1, 2\}$. All nodes in the network are assumed to have one antenna. All the channels from the base station to devices are assumed to be Rayleigh faded. The channel gain coefficient from the base station to $U_{n}$ is represented by $h_{n}$. As a result, the channel power gain, denoted by $|h_{n}|^{2}$, follows an exponential distribution having mean parameter $\lambda_{n}=L_{p}d_{n}^{-e}$, where $L_{p}$ indicates path loss constant, $d_{n}$ stands for the distance of $U_{n}$ from the base station, and $e$ refer to the path loss exponent. Without any loss of generality, we presume that the channel power gains are arranged as $|h_{1}|^{2}>|h_{2}|^{2}$. Thus, based on the channel power gain conditions, $U_{1}$ and $U_{2}$ could be referred to as strong device and weak device, respectively. The transmission power broadcasted from the base station to both devices is denoted by $P_{t}$. The fraction of transmission power $P_{t}$ allocated to $U_{1}$ is indicated by a power allocation coefficient $\alpha$, where $0\leq\alpha\leq1$. The remaining fraction $(1-\alpha)$ is allocated to  $U_2$.

\begin{Remark}
\emph{Asking all devices in the network to participate in NOMA jointly is not a good choice due to two reasons: first, sharing the same resource block among multiple devices in NOMA causes strong co-channel interference at receivers; and second, due to superposition coding and multiple SIC, long delays and high implementation complexity occur at both the transmitting and receiving ends with more devices. Therefore,  the devices of the network are divided into multiple groups, where NOMA is implemented within each group \cite{ding2016impact}, \cite{9072642}. Please note that the grouping/pairing of two devices to perform NOMA has been extensively studied in the literature in order to maintain implementation complexity and system performance.  Therefore, for the purpose of our analytical study, we consider two devices performing NOMA in one resource block, in our manuscript. However, it is possible to increase the number of devices in a group.}
\end{Remark}

\subsection{NOMA Transmission Principle}\label{NOMA_principle}
The base station first superposes all the message signals dedicated for devices and then transmits the superposed signal to each device. Thus, the broadcasted signal from the base station to both devices can be expressed as
\begin{equation}
\text{x}=\sqrt{\alpha P_{t}}\text{x}_{1} + \sqrt{(1-\alpha)P_{t}}\text{x}_{2},
\end{equation}
where $\text{x}_{1}$ and $\text{x}_{2}$, respectively, signify the message signals with unit power dedicated for $U_{1}$ and $U_{2}$. Then, the received signal $\text{y}_{n}$ for $U_{n}$, where $n \in \mathbb{N}$, can be given as
\begin{equation}
\text{y}_{n} = h_{n}\text{x} + w_{n},
\end{equation}
where $w_{n}$ represents the additive white Gaussian noise (AWGN) for $U_{n}$. Without loss of generality, we assume that $w_n$ has a mean equal to zero and variance equal to $\sigma^{2}$. 

After obtaining the superposed signal, receivers apply SIC to remove the inter-device interference imposed by the superposition and get the desired signal. During SIC, each device decodes its own and other devices' signals in a particular sequence. The collection of these sequences abide by devices is referred to as a \textit{decoding order} of the network. According to the conventional decoding order of NOMA, the strong device $U_{1}$ considers $U_2$'s signal as interference. Therefore, it decodes $U_{2} $'s signal at the first stage, applies SIC to cancel its interference, and then decodes its own signal at the second stage. Conversely, the weak device $U_{2}$ decodes its own signal at the first stage by considering the signal associated with $U_{1}$ as noise. Thus, a mutual trust is assumed between the devices that they will not intercept each other's information.

\subsection{Decoding Orders in NOMA with Untrusted Devices}\label{total_decoding_orders}
An IIoT network with untrusted devices is an antagonistic but realistic circumstance in which devices do not trust each other and always want to safeguard their data from other devices. Therefore, practically, we should consider the possibility that each device may attempt to decode the signal of itself or other device at any stage \cite{tvt_2020}, \cite{pradosh_comm_letter}, \cite{7343355}. Thus, in a two-device network, each device has two stages of decoding the signal of itself and the other device. As a result, four decoding orders exist based on the concept of permutation. We depict the $o$-th decoding order as a matrix $\mathbf{D}_{o}$ of size $2\times 2$. The $m$-th column of matrix $\mathbf{D}_{o}$ is expressed by a $2\times1$ column vector $\mathbf{d}_{m}$, which shows the sequence of SIC followed by $U_{m}$, where $m \in \mathbb{N}$. Specifically, $[\mathbf{d}_{m}]_{k}=n$ defines that signal of the device $U_{n}$ is decoded by the device $U_{m}$ at $k$-th stage, where $[\mathbf{d}_{m}]_{1} \neq [\mathbf{d}_{m}]_{2}$ and $n, k \in \mathbb{N}$. Thus, a decoding order of the network can be represented as $\mathbf{D}_{o}=[[\mathbf{d}_{1}]_{1}, [\mathbf{d}_{2}]_{1}; [\mathbf{d}_{1}]_{2}, [\mathbf{d}_{2}]_{2}]$. All decoding orders can be expressed as $\mathbf{D}_{1}=[2,2;1,1]$,  $\mathbf{D}_{2}=[2,1;1,2]$, $\mathbf{D}_{3}=[1,2;2,1]$, and $\mathbf{D}_{4}=[1,1;2,2]$. We define the set of these four decoding orders as $\mathbb{D}=\{\mathbf{D}_{o} | 1 \leq o \leq 4 \}$. 

\subsection{Achievable Data Rates and Secrecy Rates at Devices}\label{achievable_secrecyrates}
For a given decoding order $\mathbf{D}_{o}$, the data rate achieved at $U_m$ when $U_n$ is decoded by $U_m$, for all combinations of $n,m \in \mathbb{N}$ with Shannon's formula can be expressed as
\begin{equation} \label{information_rate}
R^{[o]}_{nm} = \log_{2}(1+\Gamma^{[o]}_{nm}),
\end{equation}
where $\Gamma^{[o]}_{nm}$ denotes the received signal to interference plus noise ratio (SINR) at $U_{m}$, when $U_n$ is decoded by $U_m$, and it can be given as
\begin{align}\label{SINR}
\Gamma^{[o]}_{nm} = \frac{a\hspace{0.5mm}|h_{m}|^{2}}{b\hspace{0.5mm}|h_{m}|^{2} + \frac{1}{\rho_t}},
\end{align}
where $\rho_{t}\stackrel{\Delta}{=}\frac{P_{t}}{\sigma^{2}}$ is the base station transmit signal-to-noise ratio (SNR). The parameters $a$ and $b$ required to define $\Gamma^{[o]}_{nm}$ for each combination of $m, n \in \mathbb{N}$ in all decoding orders are given in Table \ref{table1}. Note that $n=m$ in Table \ref{table1} indicates the SINR achieved by the legitimate device $U_n$ when decoding its own data, namely $\Gamma^{[o]}_{nn}$, resulting which we obtain $R^{[o]}_{nn}$ using \eqref{information_rate}. In Table \ref{table1}, $\beta_{\widehat nm}$ is the RI factor indicating the fraction of the residual error from the previous decoding stage, i.e., when $m$ imperfectly decodes $\widehat n$, where $m, \widehat n \in \mathbb{N}$ and $\widehat n \neq n$. Note that $0\leq\beta_{\widehat nm}\leq1$. Here $\beta_{\widehat nm}=0$ and $\beta_{\widehat nm}=1$, respectively, indicates perfect SIC and absolutely imperfect SIC \cite{7881111}, \cite{8755843}.

Next, in order to ensure secure communication, we utilize the concept of PLS. According to PLS, the secrecy rate of a legitimate device can be measured by the difference between the rate achieved at the legitimate device when decoding its own data and the rate achieved at another device when decoding the data of the legitimate device. Accordingly, the secrecy rate for $U_{n}$ against $U_{m}$, where $m,n\in \mathbb{N}$ can be expressed as \cite{wyner1975wire}, \cite{PLS}
\begin{equation}\label{secrecy_rate}
R^{[o]}_{sn} = \left[R^{[o]}_{nn} - R^{[o]}_{nm}\right]^{+}, \end{equation}
where $n \neq m$ and $[\Diamond]^{+} = \max\{0, \Diamond \}$.
To get a positive secrecy rate for a device, the rate of the main communication link must be greater than the rate of the eavesdropper's link, i.e., for obtaining  $R^{[o]}_{sn}>0$, the condition $R^{[o]}_{nn}> R^{[o]}_{nm}$, simplified to $\Gamma^{[o]}_{nn} > \Gamma^{[o]}_{nm}$ using \eqref{information_rate}, needs to be satisfied.  $[\Diamond]^{+} = \max\{0, \Diamond \}$ indicates that negative secrecy rates are considered to be zero.

\begin{table}[!t]
\scriptsize
	\centering
	\caption{Parameter values to define $\Gamma^{[o]}_{nm}$ for all decoding orders \cite{pradosh_comm_letter}}
	\setlength{\tabcolsep}{12pt} 
	\renewcommand{\arraystretch}{1.5} 
	\begin{tabular}{|c|c|c|c|c|}
		\hline
		$n$                & $m$                & $\mathbf{D}_{o}$                   & $a$                                 & $b$                     \\ \hline
		\multirow{4}{*}{1} & \multirow{2}{*}{1} & $\mathbf{D}_{3}$, $\mathbf{D}_{4}$ &  \multirow{4}{*}{$\alpha$}    & $(1-\alpha)$      \\ \cline{3-3}  \cline{5-5} 
		&                    & $\mathbf{D}_{1}$, $\mathbf{D}_{2}$ &                              & $(1-\alpha)\beta_{21}$ \\ \cline{2-3} \cline{5-5} 
		& \multirow{2}{*}{2} & $\mathbf{D}_{2}$, $\mathbf{D}_{4}$ &                                      & $(1-\alpha)$      \\ \cline{3-3}
		  \cline{5-5} 
		&                    & $\mathbf{D}_{1}$, $\mathbf{D}_{3}$ &                           & $(1-\alpha)\beta_{22}$  \\ \hline
		\multirow{4}{*}{2} & \multirow{2}{*}{1} & $\mathbf{D}_{3}$, $\mathbf{D}_{4}$ &  \multirow{4}{*}{$(1-\alpha)$} & $\alpha\beta_{11}$     \\ \cline{3-3}  \cline{5-5} 
		&                    & $\mathbf{D}_{1}$, $\mathbf{D}_{2}$ &                                     & $\alpha$          \\ \cline{2-3} \cline{5-5} 
		& \multirow{2}{*}{2} & $\mathbf{D}_{2}$, $\mathbf{D}_{4}$ &                                     & $\alpha\beta_{12}$     \\ \cline{3-3} \cline{5-5} 
		&                    & $\mathbf{D}_{1}$, $\mathbf{D}_{3}$ &                                     & $\alpha$          \\ \hline
	\end{tabular}
	\label{table1}
\end{table}

\section{Secure Decoding Orders}\label{section3}
As mentioned in Section \ref{total_decoding_orders}, four decoding orders are possible in the case of NOMA with two untrusted devices. Our motive is to protect each device's data from another device. Therefore, this section investigates which decoding orders are feasible in ensuring a positive secrecy rate for each device. 

\subsection{Infeasibility of Conventional Decoding Order}\label{conventional_decoding_order}
With the conventional decoding order in untrusted environment, a weak device $U_2$ may try to decode the signal of $U_1$ after cancelling the signal of itself through SIC \cite{basepaper}. Thus, the decoding order can be written as $\mathbf{D}_{1}=[2,2;1,1]$. Below in Proposition \ref{P1}, we first prove that $\mathbf{D}_{1}$ is not efficient to achieve a secure NOMA network.
\begin{Proposition}\label{P1}
\emph{With decoding order $\mathbf{D}_{1}=[2,2;1,1]$, the data of the strong device can be secured from the weak device with a constraint on power allocation, while the data of the weak device is not secured against the strong device. }
\end{Proposition}
\begin{IEEEproof}
For $\mathbf{D}_{1}=[2,2;1,1]$, the achievable SINRs $\Gamma^{[1]}_{nm}$, when $U_{m}$ decodes the signal of $U_{n}$, where $m, n \in \mathbb{N}$, with linear RI model, can be given as per Table \ref{table1} as $ \Gamma^{[1]}_{21} = \frac{(1-\alpha)|  h_{1}|^{2}}{\alpha|h_{1}|^{2} + \frac{1}{\rho_{t}} }, 
\Gamma^{[1]}_{22} = \frac{(1-\alpha)|h_{2} |^{2}}{\alpha|h_{2}|^{2} + \frac{1}{\rho_{t}} }, 
 \Gamma^{[1]}_{11} = \frac{\alpha|h_{1}|^{2}}{(1-\alpha)\beta_{21}|h_{1}|^{2} + \frac{1}{\rho_{t}}}, $ and $
\Gamma^{[1]}_{12} = \frac{\alpha|h_{2}|^{2}}{(1-\alpha)\beta_{22}|h_{2}|^2 + \frac{1}{\rho_{t}}}$. To get positive secrecy rate for $U_{1}$, we solve the required condition $\Gamma^{[1]}_{11} > \Gamma^{[1]}_{12}$ as explained in Section \ref{achievable_secrecyrates} and get a feasible condition on $\alpha$ as
\begin{equation}
\alpha < 1+\frac{|h_{1}|^2 - |h_{2}|^2 }{|h_{1}|^2 |h_{2}|^2 \rho_t (\beta_{22}-\beta_{21})}.
\end{equation}
 Note that $\alpha=0$ gives $R^{[1]}_{s1}=0$, and hence, infeasible. As a result,  a positive secrecy rate can be obtained for strong device $U_1$ against $U_2$ with a constraint on power allocation as
\begin{equation}
0<\alpha<1+\frac{|h_{1}|^2 - |h_{2}|^2 }{|h_{1}|^2 |h_{2}|^2 \rho_t (\beta_{22}-\beta_{21})}.
\end{equation}
On the other hand, the condition $\Gamma^{[1]}_{22} > \Gamma^{[1]}_{21}$ to get positive secrecy rate for $U_{2}$ gives $|h_{2}|^{2} >| h_{1}|^{2}$, which is infeasible since we assume that $|h_{1}|^{2} >| h_{2}|^{2}$ (Refer Section \ref{network_model}). Hence, we cannot obtain a positive secrecy rate for $U_{2}$.
\end{IEEEproof}
 
Thus, it can be concluded that $\mathbf{D}_{1}$ is not a feasible decoding order in achieving a positive secrecy rate to both devices.

\subsection{Feasibility Check for Other Possible Decoding Orders}\label{secure_decoding_orders}
Now we check the feasibility of decoding orders $\mathbf{D}_{2}$, $\mathbf{D}_{3}$, and $\mathbf{D}_{4}$, one by one, in achieving secure NOMA transmission. 
\subsubsection{Feasibility Check for $\mathbf{D}_{2}=[2,1;1,2]$ }\label{feasibility_check_D2}
A key result on the feasibility of $\mathbf{D}_{2}$ is provided below in Proposition \ref{P2}.
\begin{Proposition}\label{P2}
\emph{The decoding order $\mathbf{D}_{2}=[2,1;1,2]$ is feasible in achieving a secure NOMA communication in untrusted scenario as we can get a positive secrecy rate for both devices with a constraint on power allocation as
\begin{equation}\label{alpha_bound_D2}
 \frac{|h_{1}|^{2}-|h_{2}|^{2}}{|h_{1}|^{2}|h_{2}|^{2}\rho_{t}(1-\beta_{12})} < \alpha < 1.
\end{equation}}
\end{Proposition}
\begin{IEEEproof}
According to $\mathbf{D}_{2}=[2,1;1,2]$, each device first decodes the signal of other device, and then decodes its own signal after performing SIC. As a result, the received SINRs as per Table \ref{table1} are given as $
\Gamma^{[2]}_{21} = \frac{(1-\alpha)|h_{1}|^{2}}{\alpha|h_{1}|^{2}+\frac{1}{\rho_{t}}},  \Gamma^{[2]}_{12} = \frac{\alpha|h_{2}|^{2}}{(1-\alpha)|h_{2}|^{2} + \frac{1}{\rho_{t}}}, \Gamma^{[2]}_{11} = \frac{\alpha|h_{1}|^{2}}{(1-\alpha)\beta_{21}|h_{1}|^{2}+\frac{1}{\rho_{t}}},$ and $
\Gamma^{[2]}_{22} = \frac{(1-\alpha)|h_{2}|^{2}}{\alpha\beta_{12}|h_{2}|^{2}+\frac{1}{\rho_{t}}}.$ The condition $\Gamma^{[2]}_{11} > \Gamma^{[2]}_{12}$ to obtain positive secrecy rate $R^{[2]}_{s1}$ for $U_{1}$ leads to a feasible condition as
\begin{equation}\label{alpha_bound_1_D2}
\alpha < 1 + \frac{|h_{1}|^{2}-|h_{2}|^{2}}{|h_{1}|^{2}|h_{2}|^{2}\rho_{t}(1-\beta_{21})}.
\end{equation}
Note that $0\leq\alpha\leq1$ (Refer Section \ref{network_model}) and $0\leq\beta_{21}\leq1$ (Refer Section \ref{achievable_secrecyrates}). Also, $\alpha=0$ gives $R^{[2]}_{s1}=0$. Hence, the condition on power allocation to get a positive secrecy rate for the strong device $U_1$ will be $0<\alpha\leq1$.

Similarly, $\Gamma^{[2]}_{22}>\Gamma^{[2]}_{21}$ for $R^{[2]}_{s2}>0$ gives a condition as
\begin{equation}\label{alpha_bound_2_D2}
\alpha >  \frac{|h_{1}|^{2}-|h_{2}|^{2}}{|h_{1}|^{2}|h_{2}|^{2}\rho_{t}(1-\beta_{12})}.
\end{equation}
Here also due to three conditions, i.e., $0\leq\alpha\leq1$ (Refer Section \ref{network_model}), $0\leq\beta_{12}\leq1$ (Refer Section \ref{achievable_secrecyrates}), and $R^{[2]}_{s2}=0$ with $\alpha=1$,  the feasible power allocation condition for $R^{[2]}_{s2}>0$ will be $\frac{|h_{1}|^{2}-|h_{2}|^{2}}{|h_{1}|^{2}|h_{2}|^{2}\rho_{t}(1-\beta_{12})}<\alpha<1$. 

From the analysis mentioned above, we see that a positive secrecy rate can be acquired for each device against the other device in an untrusted scenario if decoding orders $\mathbf{D}_{2}$ is followed, with suitable power allocation constraints. Thus, jointly, the power allocation condition providing secrecy to both devices can be written as given in \eqref{alpha_bound_D2}.
\end{IEEEproof}

Similar to the above analysis, we can check the feasibility of $\mathbf{D}_{3}$ and $\mathbf{D}_{4}$ in achieving a secure NOMA network.

\subsubsection{Feasibility Check of $\mathbf{D}_{3}=[1,2;2,1]$ }\label{feasibility_check_D3}
 Below a key result on feasibility of $\mathbf{D}_{3}$ is provided in Proposition 3.
\begin{Proposition}\label{P3}
\emph{Decoding order  $\mathbf{D}_{3}=[1,2;2,1]$ is infeasible for achieving secure NOMA communication among untrusted devices as only the data of strong device can be secured from the weak device with a constraint on power allocation as 
\begin{equation}\label{alpha_bound_D3}
\alpha>1-\frac{|h_{1}|^{2}-|h_{2}|^{2}}{|h_{1}|^{2}|h_{2}|^{2}\rho_{t}(1-\beta_{22})}.
\end{equation}}
\end{Proposition}

\begin{IEEEproof}
The proof is given in Appendix A.
\end{IEEEproof}

\subsubsection{Feasibility Check for $\mathbf{D}_{4}=[1,1;2,2]$}\label{feasibility_check_D4}
Now we present a key result on the feasibility of $\mathbf{D}_{4}$ via Proposition \ref{P4}.
\begin{Proposition}\label{P4}
\emph{The decoding order $\mathbf{D}_{4}=[1,1;2,2]$ is a feasible secure decoding order because it is efficient in providing positive secrecy rate for both devices in an untrusted NOMA network with a constraint on power allocation as
\begin{equation}\label{alpha_bound_D4}
 \frac{|h_{1}|^{2}-|h_{2}|^{2}}{|h_{1}|^{2}|h_{2}|^{2}\rho_{t}(\beta_{11}-\beta_{12})} < \alpha < 1.
\end{equation}}
\end{Proposition}
\begin{IEEEproof}
The proof is provided in Appendix B.
\end{IEEEproof}
\begin{Remark}
\emph{In the case of $\mathbf{D}_{4}=[1,1;2,2]$, there exists a special condition, i.e., when  $\beta_{11}\leq\beta_{12}$, then on solving $\Gamma^{[4]}_{22}>\Gamma^{[4]}_{21}$, we find an infeasible condition, which shows that a positive secrecy rate cannot be acquired for the weak device. Thus, if $\beta_{11}>\beta_{12}$, only then  $\mathbf{D}_{4}$ is a feasible decoding order for secure NOMA transmission.}
\end{Remark}
\begin{Remark}\label{remark3}
\emph{From Propositions \ref{P1}-\ref{P4}, we observe that a positive secrecy rate can be received for each device with a suitable constraint on power allocation in decoding orders $\mathbf{D}_{2}$ and $\mathbf{D}_{4}$. Therefore, we refer to $\mathbf{D}_{2}$ and $\mathbf{D}_{4}$ as secure decoding orders. For further analysis, we define the set of these two secure decoding orders as $\mathbb{S}=\{\mathbf{D}_{o} | o \in 2, 4 \}$.}
\end{Remark}

\section{Secrecy Fairness Maximization}\label{section4}
 In this section, we aim to optimize resources, such as decoding order and power allocation, from the perspective of secrecy fairness. A secrecy fairness viewpoint means that network resources should be allocated to devices in such a way that the secrecy rate performance of each device is ensured. Note that since NOMA primarily pairs or groups devices with significantly different channel gains, the fundamental basis for examining secrecy fairness is that weak devices can also obtain sufficient communication resources in the same manner as strong devices so no loss of secrecy rate performance occurs for weak devices. Under the secrecy fairness criterion, we focus on maximizing the minimum secrecy rate between devices. In this regard, we first formulate the optimization problem, then provide a solution methodology, followed by its closed-from optimal solution.

\subsection{Problem Formulation}
As we obtained in Section \ref{section3} that there are two secure decoding orders $\mathbf{D}_{2}$ and $\mathbf{D}_{4}$ that can ensure a positive secrecy rate for each device, we will optimize the decoding order over set $\mathbb{S}$ of secure decoding orders (Refer Remark \ref{remark3}). Also, since $R^{[o]}_{s1}$ and $R^{[o]}_{s2}$, is a function of both decoding order and $\alpha$, we formulate a joint optimization problem as maximizing the minimum secrecy rate between devices over a set $\mathbb{S}$ of secure decoding orders and power allocation coefficient $\alpha$ as
\begin{align}
\mathcal{P}_1: & \quad \underset{\mathbf{D}_{o}\in \mathbb{S}, \alpha}{{\max}} \quad  \min\left[R^{[o]}_{s1}, R^{[o]}_{s2}\right], \nonumber
\\  \text{s.t.} & \quad \mathcal{C}_1: 0 \leq \alpha \leq 1,  \quad \mathcal{C}_2: R^{[o]}_{s1}>0, \quad \mathcal{C}_3: R^{[o]}_{s2}>0, \nonumber
\end{align}
where $\mathcal{C}_1$ refers to the constraint on power allocation coefficient (Refer Section \ref{network_model}), and $\mathcal{C}_{2}$ and $\mathcal{C}_{3}$ denote the positive secrecy rate conditions for $U_1$ and $U_2$, respectively.

\subsection{Solution Methodology}\label{solution_methodology}
We observe that $\mathcal{P}_1$ is a combinatorial optimization problem because there are two secure decoding orders, and the secrecy rate depends on $\alpha$ in each secure decoding order. Therefore, to reduce the computational complexity in determining the joint optimal solution of decoding order and power allocation, we solve the joint optimization problem $\mathcal{P}_1$ in two steps as described below.
\begin{itemize}
    \item In the first step, ignoring the power allocation constraint, we optimize the decoding order over a set $\mathbb{S}$ of secure decoding orders for maximizing the minimum secrecy rate between the devices. This way we find the optimal secure decoding order for maximizing the minimum secrecy rate between devices. (Refer Section \ref{section_optimal_secure_decoding_order}) 
    \item In the second step, to complete joint optimization, we obtain the optimal power allocation solution in closed form for only optimal secure decoding order obtained in the first step. (Refer Section \ref{section_optimal_power_allocation})
\end{itemize} 

\subsection{Solution of Joint Optimization Problem}
\subsubsection{Optimal Secure Decoding Order}\label{section_optimal_secure_decoding_order}
Following the solution methodology's first step, an investigation of optimal secure decoding order is presented through Lemma \ref{optimal_secure_decoding_order}.
\begin{Lemma}\label{optimal_secure_decoding_order}
\emph{The optimal secure decoding order that maximizes the minimum secrecy rate between devices is $\mathbf{D}_{\widehat o}=\mathbf{D}_{2}$, where $\widehat o=2$ is the index of optimal secure decoding order.}
\end{Lemma}

\begin{IEEEproof}
Here, considering the set $\mathbb{S}$ of secure decoding orders, we compare secrecy rates obtained with decoding orders $\mathbf{D}_{2}$ and $\mathbf{D}_{4}$. The secrecy rate is in the form of $\log2\left(\frac{1+\frac{Q}{R}}{1+\frac{S}{T}}\right)$. In case of $\mathbf{D}_{2}$, $\mathrm{R}=(1-\alpha)\beta_{21}|h_{1}|^{2}+\frac{1}{\rho_{t}}$ for $R^{[2]}_{s1}$, and $\mathrm{T}=\alpha|h_{1}|^{2}+\frac{1}{\rho_{t}}$ for $R^{[2]}_{s2}$. Coming to $\mathbf{D}_{4}$, we have $\mathrm{R}=(1-\alpha)|h_{1}|^{2}+\frac{1}{\rho_{t}}$ for $R^{[4]}_{s1}$, and  $\mathrm{T}=\alpha\beta_{11}|h_{1}|^{2}+\frac{1}{\rho_{t}}$ for $R^{[4]}_{s2}$. The other parameters are the same in both secure decoding orders. 
Here, in the case of secrecy rate for $U_{1}$, $\mathrm{R}$ is lower in $\mathbf{D}_{2}$ than in $\mathbf{D}_{4}$ since $\beta_{21}<1$. Thus, $\mathbf{D}_{2}$ ensures more secrecy rate for $U_{1}$ than $\mathbf{D}_{4}$. Similarly, we observe that in the case of $U_{2}$, $\mathbf{D}_{2}$ again ensures more secrecy rate for $U_{2}$ than $\mathbf{D}_{4}$ since $\mathrm{T}$ is higher in $\mathbf{D}_{2}$ as compared to $\mathbf{D}_{4}$, since $\beta_{11}<1$. Thus, since $\mathbf{D}_{2}$ ensures more secrecy for each device than $\mathbf{D}_{4}$, it will be optimal for maximizing the minimum secrecy rate between devices. Hence, $\mathbf{D}_{2}$ is an optimal secure decoding order for secrecy fairness maximization between devices.
\end{IEEEproof}

\subsubsection{Optimal Power Allocation}\label{section_optimal_power_allocation}
We already have solved $\mathcal{P}_1$ over the set $\mathbb{S}$ of secure decoding orders by obtaining the optimal secure decoding order $\mathbf{D}_{\widehat o}$ in Section \ref{section_optimal_secure_decoding_order}. Therefore, to complete the joint optimization, we can now solve $\mathcal{P}_1$ over $\alpha$ for only $\mathbf{D}_{\widehat o}$. In this regard, considering $\widehat o$ as the index of optimal secure decoding order, $\mathcal{P}_1$ can be restated as
\begin{align}
\mathcal{P}_{1a}: & \quad \underset{\alpha}{\max} \quad  \min\left[R^{[\widehat o]}_{s1}, R^{[\widehat o]}_{s2}\right], \nonumber
\\  \text{s.t.} & \quad \mathcal{C}_{1},  \quad \mathcal{C}_{4}: R^{[\widehat o]}_{s1}>0, \quad \mathcal{C}_{5}: R^{[\widehat o]}_{s2}>0, \nonumber
\end{align}
where $\mathcal{C}_{4}$ and $\mathcal{C}_{5}$ are positive secrecy rate conditions for $U_{1}$ and $U_{2}$, respectively, in optimal secure decoding order $\mathbf{D}_{\widehat o}$. 

Following Lemma \ref{optimal_secure_decoding_order}, $\widehat o=2$. Thus, $\mathcal{P}_{1a}$ can be restated as
\begin{align}
\mathcal{P}_{1b}: & \quad \underset{\alpha}{\max} \quad  \min\left[R^{[2]}_{s1}, R^{[2]}_{s2}\right], \nonumber
\\  \text{s.t.} & \quad \mathcal{C}_{6}: \frac{|h_{1}|^{2}-|h_{2}|^{2}}{|h_{1}|^{2}|h_{2}|^{2}\rho_{t}(1-\beta_{12})} < \alpha < 1, \nonumber
\end{align}
where $\mathcal{C}_{6}$ refers to the constraint on power allocation coefficient $\alpha$, which is obtained in \eqref{alpha_bound_D2} by solving $\mathcal{C}_{1}$, $\mathcal{C}_{4}$, and $\mathcal{C}_{5}$ for $\widehat o=2$ (Refer Proposition \ref{P2} in Section \ref{feasibility_check_D2}).

Further, using $x_c=\min\left[R^{[2]}_{s1}, R^{[2]}_{s2}\right]$, $\mathcal{P}_{1b}$ can be written as
\begin{align}\label{problem_re_formulation}
\mathcal{P}_{1c}: & \quad \underset{\alpha, x_c}{\max} \quad  x_{c}, \nonumber
\\  \text{s.t.} & \quad \mathcal{C}_{6}, \quad \mathcal{C}_{7}: x_{c}\leq R^{[2]}_{s1}, \quad \mathcal{C}_{8}: x_{c}\leq R^{[2]}_{s2}, \nonumber
\end{align}
where $ \mathcal{C}_{7}$ and $ \mathcal{C}_{8}$ comes from the definition of $\min[.]$.

Note that $\mathcal{P}_{1c}$ is a non-convex problem because of the presence of non-convex constraints $\mathcal{C}_{7}$ and $\mathcal{C}_{8}$. That is why finding the optimal solution of power allocation is challenging. Therefore, we solve the optimization problem $\mathcal{P}_{1c}$ by obtaining all possible optimal points with KKT conditions,  which are the candidates for the global-optimal solution  \cite{ravindran2006engineering}. After obtaining candidate optimal points, we can select the global-optimal solution as the feasible optimal point that maximizes the minimum secrecy rate between devices. The global-optimal power allocation solution of  $\mathcal{P}_{1c}$ is given by Lemma \ref{Lemma_1}.
\begin{Lemma}\label{Lemma_1}
\textit{The global-optimal power allocation solution, denoted by $\widehat \alpha$, of $\mathcal{P}_{1c}$, is the feasible candidate from the obtained candidates that maximizes the minimum secrecy rate between devices and can be  given  as
\begin{equation}\label{optimal_a_asy_sop}
\widehat \alpha \stackrel{\Delta}{=}\!\underset{\alpha \in \{\alpha_{2}^{*}, \alpha_{3}^{*}, \alpha_{4}^{*}, \alpha_{5}^{*}, \alpha_{6}^{*}, \alpha_{7}^{*}\}}{\mathrm{argmax}}\! \min\left[R^{[2]}_{s1},R^{[2]}_{s2}\right],
\end{equation}
where $\alpha_{2}^{*}, \alpha_{3}^{*}, \alpha_{4}^{*}, \alpha_{5}^{*}, \alpha_{6}^{*}, \alpha_{7}^{*}$ are the candidate optimal points and each of them is obtained in the closed-form as described in the proof.}
\end{Lemma}

\begin{IEEEproof}
To solve $\mathcal{P}_{1c}$, we keep the boundary constraint on power allocation coefficient, i.e., $\mathcal{C}_{6}$, implicit and connect Lagrange multipliers $\delta_{1}$ with $\mathcal{C}_{7}$ and $\delta_{2}$ with $\mathcal{C}_{8}$. Thus, we can define the Lagrangian function $L$ as
\begin{equation}\label{lagragian_function1}
    L=x_{c}-\delta_{1}\left[x_{c}-R^{[2]}_{s1}\right]-\delta_{2}\left[x_{c}-R^{[2]}_{s2}\right].
\end{equation}
There are 4 KKT conditions. The primal feasibility conditions are given by $\mathcal{C}_{7}$ and $\mathcal{C}_{8}$. The dual feasibility conditions are $\delta_{1}\geq 0$ and $\delta_{2}\geq 0$. The subgradient conditions are given as
\begin{subequations}
\begin{align}
\frac{\mathrm{d}L}{\mathrm{d} x_{c}}  &=  1-\delta_{1}-\delta_{2}= 0, \label{subgradient_1}\\
\frac{\mathrm{d}L}{\mathrm{d}\alpha}  &=  \delta_{1} \frac{\mathrm{d} R^{[2]}_{s1}}{\mathrm{d}\alpha}  + \delta_{2}\frac{\mathrm{d} R^{[2]}_{s2}}{\mathrm{d}\alpha} = 0. \label{subgradient_2}
\end{align}
\end{subequations}
The two complementary slackness conditions are expressed as
\begin{subequations}
\begin{align}
 \delta_{1} \left[ R^{[2]}_{s1} - x_{c}\right] = 0, \label{complementary_slackness_1} \\
 \delta_{2} \left[ R^{[2]}_{s2} -  x_{c}\right] = 0. \label{complementary_slackness_2}
\end{align}
\end{subequations}

Each of the Lagrange multipliers, i.e., $\delta_{1}$ and $\delta_{2}$, could be either equal to or greater than zero. Thus, 4 cases exist, which are discussed in the following.

\emph{Case 1: $\delta_{1} = 0$, $\delta_{2}=  0$}: This implies $\delta_{1} + \delta_{2} = 0$. However, as given in \eqref{subgradient_1}, $\delta_{1} +  \delta_{2} = 1$. Thus, this is an infeasible case.

\emph{Case 2: $\delta_{1} = 0$, $\delta_{2} > 0$}: This case, using \eqref{subgradient_2}, implies $\frac{\mathrm{d}L}{\mathrm{d}\alpha} = \frac{\mathrm{d} R^{[2]}_{s2}}{\mathrm{d}\alpha} = 0$. 
On solving $\frac{\mathrm{d}R^{[2]}_{s2}}{\mathrm{d}\alpha}\!=\!0$, we obtain a quadratic equation solving which on $\alpha$, we get two roots, denoted by $\alpha^{*}_{1}$ and $\alpha^{*}_{2}$, as given in (17) on the top of next page. We observe that $\!\alpha^{*}_{1}\!=\!\dfrac{\!\left(1\!-\!\beta_{12}\right)\!|h_{1}|^{2}\!\!+\!\!\sqrt{\!\left(1\!-\!\beta_{12}\right)\! |h_{1}|^{2} (|h_{1}|^{2}\!\!-\!\!\beta_{12}|h_{2}|^{2})(\beta_{12}|h_{2}|^{2}\rho_{t}\!+\!1)}}{\beta_{12}\left(\beta_{12}\!-\!1\right)|h_{1}|^{2}|h_{2}|^{2}\rho_t}\!$ is infeasible. The reason is that for $(|h_{1}|^{2}-\beta_{12}|h_{2}|^{2})>0$, the required condition is $\beta_{12}<\frac{|h_{1}|^{2}}{|h_{2}|^{2}}$, which is true since $\beta_{12}<1$ and $|h_{1}|^{2}>|h_{2}|^{2}$. As a result, $\alpha^{*}_{1}$ is negative, which is infeasible. Therefore, we consider the root $\alpha_{2}^{*}$ as the candidate for the optimal power allocation solution.

\begin{figure*}
\begin{equation}\label{case2_roots}
\alpha^{*}_{1},\alpha^{*}_{2}=\dfrac{\left(1-\beta_{12}\right)|h_{1}|^{2}\pm\sqrt{\left(1-\beta_{12}\right) |h_{1}|^{2} (|h_{1}|^{2}-\beta_{12}|h_{2}|^{2})(\beta_{12}|h_{2}|^{2}\rho_{t}+1)}}{\beta_{12}\left(\beta_{12}-1\right)|h_{1}|^{2}|h_{2}|^{2}\rho_t},
\end{equation}
\noindent\rule{18cm}{0.5pt}
\end{figure*}

\emph{Case 3: $\delta_{1} > 0$, $\delta_{2} = 0$}:
In the third case, using \eqref{subgradient_2},  $\frac{\mathrm{d}L}{\mathrm{d}\alpha} = \frac{\mathrm{d} R^{[2]}_{s1}}{\mathrm{d}\alpha} = 0$ is obtained.
Similar to the case 2, here also $\frac{\mathrm{d}R^{[2]}_{s1}}{\mathrm{d}\alpha}=0$ leads to a quadratic equation in terms of $\alpha$ which gives two roots $\alpha^{*}_{3}$ and $\alpha^{*}_{4}$, as given in \eqref{case3_roots} on the next page. These roots also are the candidates for the optimal solution.

\begin{figure*}
\begin{equation}\label{case3_roots}
\alpha^{*}_{3},\alpha^{*}_{4}=    \dfrac{\left(\beta_{21}-1\right)|h_{2}|^{2}(\beta_{21}|h_{1}|^{2}\rho_t+1)\!\pm\!\sqrt{(1-\beta_{21})|h_{2}|^{2} (\beta_{21}|h_{1}|^{2}\rho_t+1) (|h_{2}|^{2} - \beta_{21}|h_{1}|^{2}) }}{\beta_{21}\left(\beta_{21}-1\right)|h_{1}|^{2}|h_{2}|^{2}\rho_t},
\end{equation}
\noindent\rule{18cm}{0.5pt}
\end{figure*}

\emph{Case 4: $\delta_{1} > 0$, $\delta_{2} > 0$}: Using \eqref{complementary_slackness_1} and \eqref{complementary_slackness_2}, this case implies $R^{[2]}_{s1}=R^{[2]}_{s2}$, which indicates equal secrecy rate for both devices. Thus, using \eqref{information_rate}, \eqref{SINR}, \eqref{secrecy_rate}, and Table \ref{table1}, we solve $R^{[2]}_{s1}=R^{[2]}_{s2}$ for decoding order $\mathbf{D}_{2}$, which can be given as
\begin{align}\label{Rs1_equal_Rs2}
    \log_{2}\frac{\Big(1+\frac{\alpha \hspace{0.5mm}|h_{1}|^{2}}{(1-\alpha)\beta_{21}\hspace{0.5mm}|h_{1}|^{2} + \frac{1}{\rho_t}}\Big)}{\Big(1+ \frac{\alpha \hspace{0.5mm}|h_{2}|^{2}}{(1-\alpha)\hspace{0.5mm}|h_{2}|^{2} + \frac{1}{\rho_t}}\Big)}   &= \log_{2}\frac{\Big(1+ \frac{(1-\alpha) \hspace{0.5mm}|h_{2}|^{2}}{\alpha\beta_{12}\hspace{0.5mm}|h_{2}|^{2} + \frac{1}{\rho_t}}\Big)}{\Big(1+ \frac{(1-\alpha) \hspace{0.5mm}|h_{1}|^{2}}{\alpha\hspace{0.5mm}|h_{1}|^{2} + \frac{1}{\rho_t}}\Big)}.
\end{align}
After some algebric simplifications in \eqref{Rs1_equal_Rs2}, a cubic equation in the form of $M_{1}\alpha^3+M_2\alpha^2+M_3\alpha+M_4=0$ is resulted with coefficients $M_1, M_2, M_3$ and $M_4$, where $M_1=B_1E_1G_1I_1-F_1H_1C_1D_1$, $M_2=B_1E_1I_1+(A_1E_1+B_1D_1)G_1I_1-F_1H_1A_1D_1-(D_1H_1+F_1)C_1D_1$, $M_3=(A_1E_1+B_1D_1)I_1+A_1D_1G_1I_1-(D_1H_1+F_1)A_1D_1-C_1D_1^2$, and $M_4=A_1D_1I_1-A_1D_1^2$ with $A_1=\beta_{21}|h_{1}|^{2}\rho_t+1$, $B_1=(|h_{1}|^{2}-\beta_{21}|h_{1}|^{2})\rho_t$, $C_1=-\beta_{21}|h_{1}|^{2}\rho_t$, $D_1=|h_{2}|^{2}\rho_t+1$, $E_1=-|h_{2}|^{2}\rho_t$, $F_1=(\beta_{12}-1)|h_{2}|^{2}\rho_t$, $G_1=\beta_{12}|h_{2}|^{2}\rho_t$, $H_1=|h_{1}|^{2}\rho_t$ , and $I_1=|h_{1}|^{2}\rho_t+1$. Thus, in this case, three roots exist, i.e., candidate optimal points denoted as $\alpha^{*}_{5}$, $\alpha^{*}_{6}$ and $\alpha^{*}_{7}$. 

The above analysis shows six candidates for optimal solution. Therefore, the global-optimal power allocation solution $\widehat \alpha$ of $\mathcal{P}_{1c}$ is the feasible candidate for which the minimum secrecy rate between the strong and weak devices is maximum, as given in \eqref{optimal_a_asy_sop}.
\end{IEEEproof}

\section{Numerical Results}\label{section5}
This section provides numerical results to validate the derived results and present key insights on the optimized solution. The default network parameters are considered as: $d_{1}=50$, $d_{2}=100$, $L_{p}=1$, and $e=3$.
Small-scale fading is supposed to follow an exponential distribution having a mean value equal to 1 at each link \cite{basepaper}. We average the simulations over $10^3$ randomly generated channel gain realizations with Rayleigh distribution for each link. Simulation and analytical results, respectively, are marked as `Sim' and `Ana'.

\begin{figure}[!t]
\centering
\includegraphics[scale=.3]{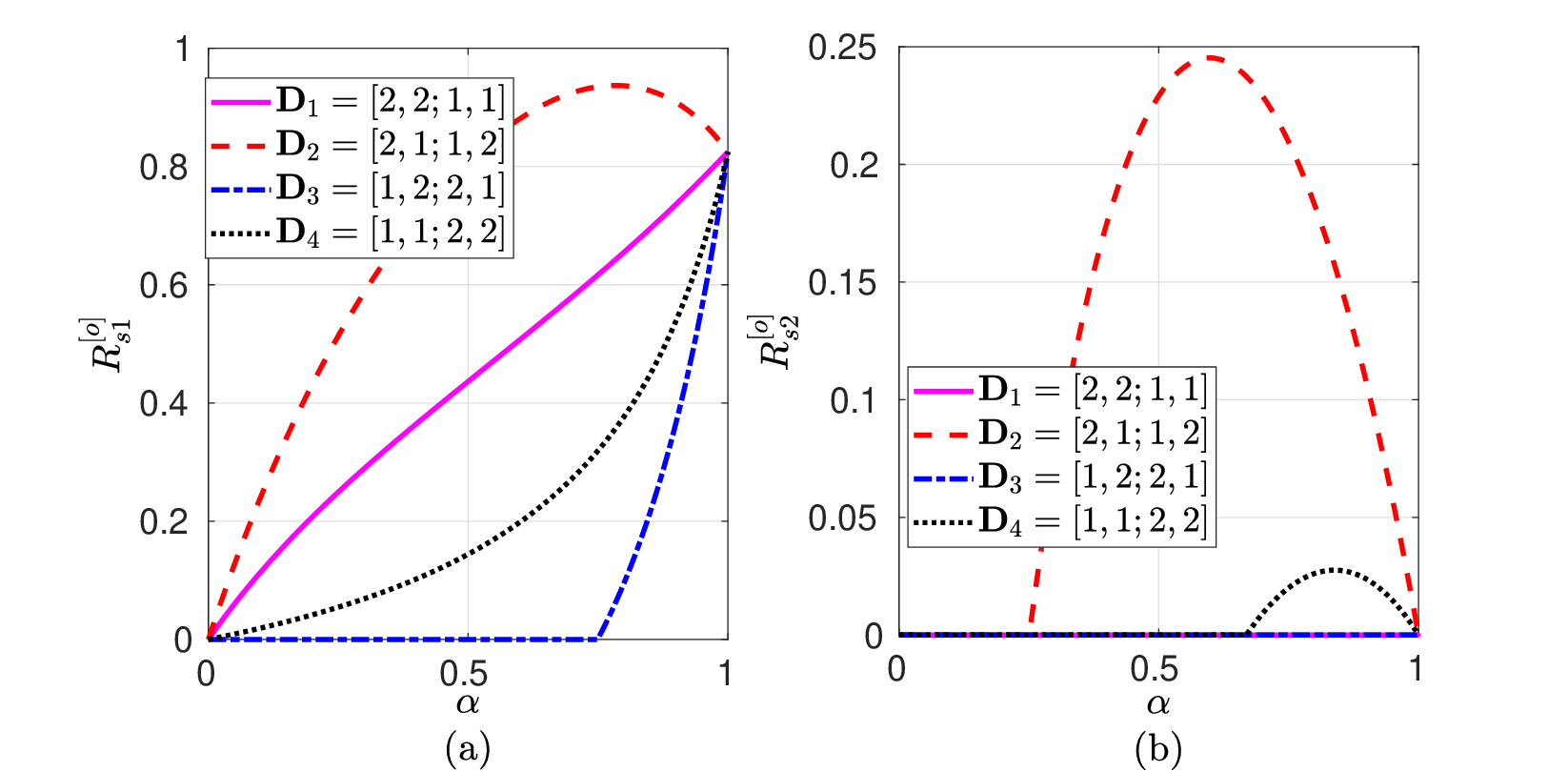}
\caption{Validation of the optimality of decoding order with $\beta_{21}=0.2$, $\beta_{22}=0.2$, $\beta_{12}=0.2$, $\beta_{11}=0.5$, and $\rho_{t}=60$ dB.}
\label{optimal_decoding_order}
\end{figure}

\begin{figure}[!t]
\centering
\includegraphics[scale=.3]{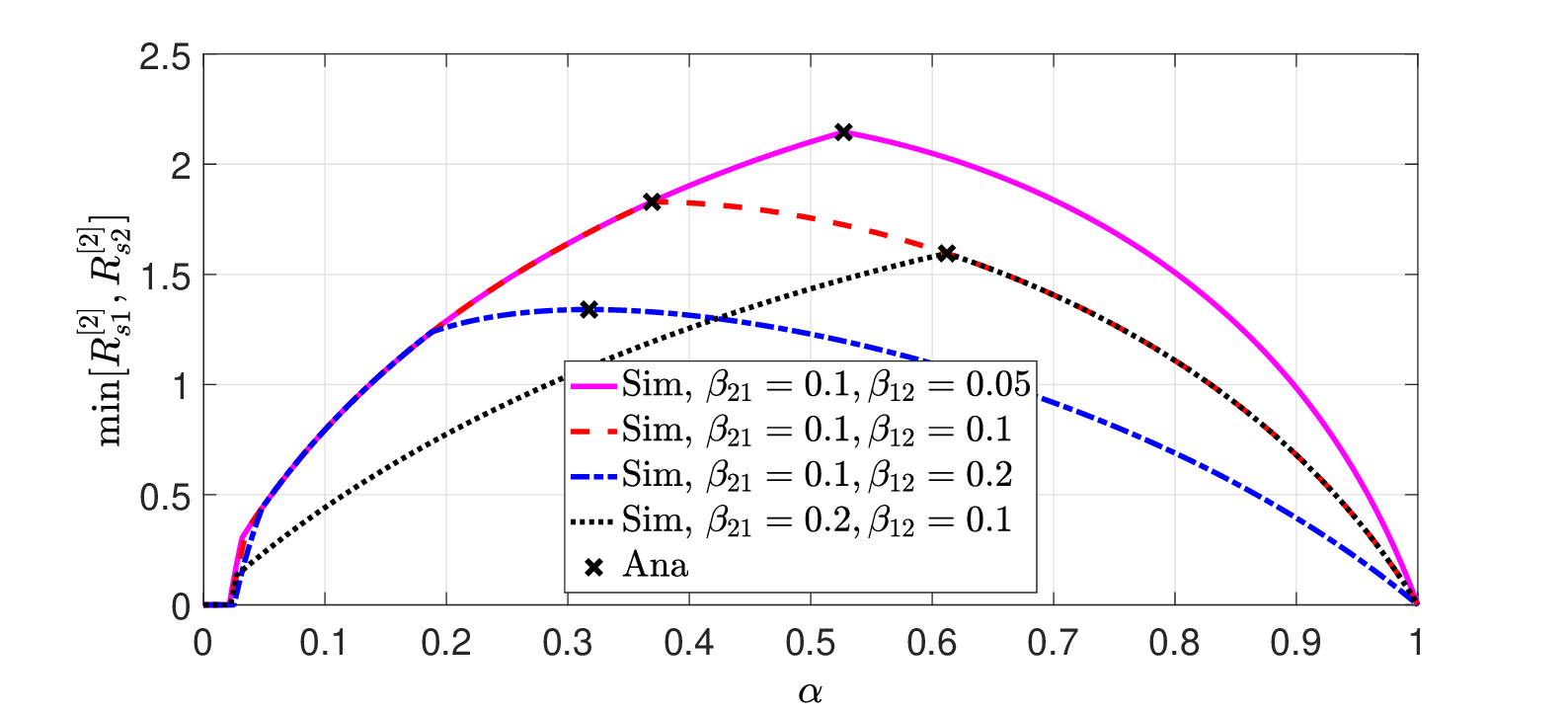}
\caption{Validation of the correctness of the closed-form optimal power allocation solution with different values of RI factor and $\rho_{t}=70$ dB.}
\label{optimal_power_allocation}
\end{figure}

\begin{figure}[!t]
\centering
\includegraphics[scale=.3]{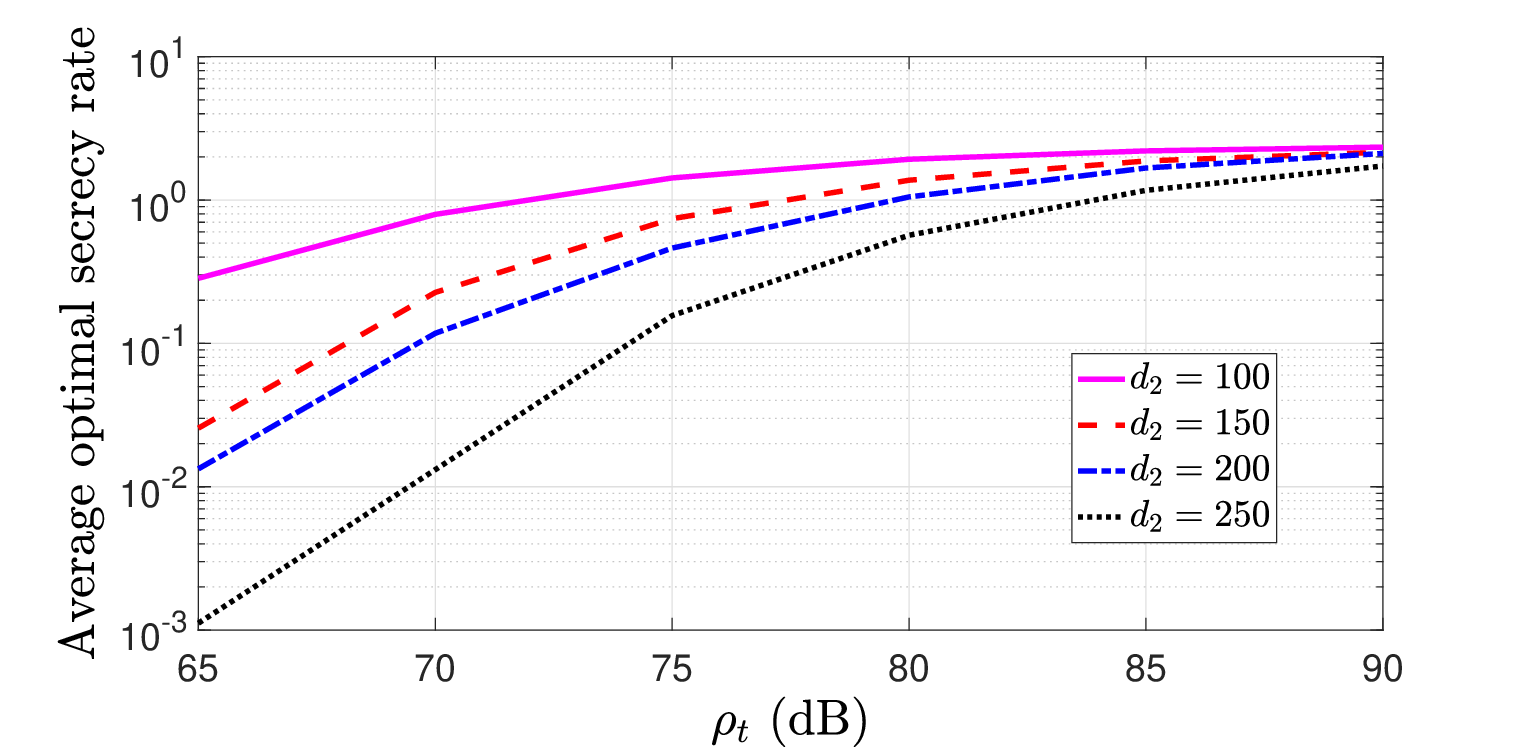}
\caption{Variation in average optimal secrecy rate performance obtained by solving the formulated max-min joint optimization problem with $\rho_t$ for different values of far device's distance $d_2$ from the base station, $\beta_{21}=\beta_{12}=0.2$.}
\label{optimal_performance_with_parameters}
\end{figure}

\subsection{Validation of Optimal Results}
Firstly, Fig. \ref{optimal_decoding_order} is plotted to validate the accuracy of Lemma 1, stating that the optimal secure decoding order maximizing the minimum secrecy rate between devices is $\mathbf{D}_{2}$. Here, through Fig. \ref{optimal_decoding_order}(a) and  Fig. \ref{optimal_decoding_order}(b), the variation in secrecy rates $R_{s1}^{[o]}$ and $R_{s2}^{[o]}$ for $U_1$ and $U_2$, respectively, with $\alpha$ for all four decoding orders is shown. The results confirm that for all $\alpha$ values, $\mathbf{D}_{2}$ provides more secrecy rate for each device than other decoding orders. Hence, to maximize the minimum secrecy rate, the optimal secure decoding order is $\mathbf{D}_{2}$.

Further, to validate optimal power allocation solution (Refer Lemma 2) for optimal secure decoding order $\mathbf{D}_{2}$, Fig. \ref{optimal_power_allocation} is plotted. Here, we show the variation of $\min\left[R_{s1}^{[2]}, R_{s2}^{[2]}\right]$ with $\alpha$ for different values of $\beta_{21}$ and $\beta_{12}$. Results indicate that there exists a unique global-optimal solution for $\min\left[R_{s1}^{[2]}, R_{s2}^{[2]}\right]$ in terms of $\alpha$. The perfect match between the simulation and analytical results confirms the accuracy of the analysis. We also observe from the results that the optimal power allocation can be greater than 0.5. It means providing lesser power to the strong device than the power allocated to the weak device, as presumed in many works in the NOMA literature, is not always necessary. Thus, we conclude that the power allocation associated with devices in a NOMA-enabled IIoT network should be decided based on the given network parameters.

\subsection{Impact of Network Parameters on Optimal Solution}
Through the results presented in Fig.  \ref{optimal_performance_with_parameters}, we study the impact of $\rho_t$ and different values of far device's distance $d_2$ on the average optimal secrecy rate performance of the network. $d_{1}$ is set to as $50$ meters. We observe that average optimal secrecy rate increases by increasing $\rho_t$. The reason is that the achievable data rates for devices increase with an increase in SNR. Here we also notice that the average optimal secrecy rate performance decreases with an increase in distance $d_{2}$. The reason is that increasing the distance $d_2$ results in a decrease in the achievable data rate of $U_2$, and consequently, an increase in secrecy rate for $U_1$ and a decrease in secrecy rate for $U_2$ is obtained. Through simulations, we notice that less secrecy rate is obtained for $U_2$ for most channel realizations. Therefore,  while calculating the average max min secrecy rate, the secrecy rate for $U_2$ is dominant for given network parameters, which decreases by increasing $d_2$. Therefore, the results show that average performance degrades with an increase in $d_2$.

\subsection{Performance Comparison}
\begin{figure}[!t]
\centering
\includegraphics[scale=.3]{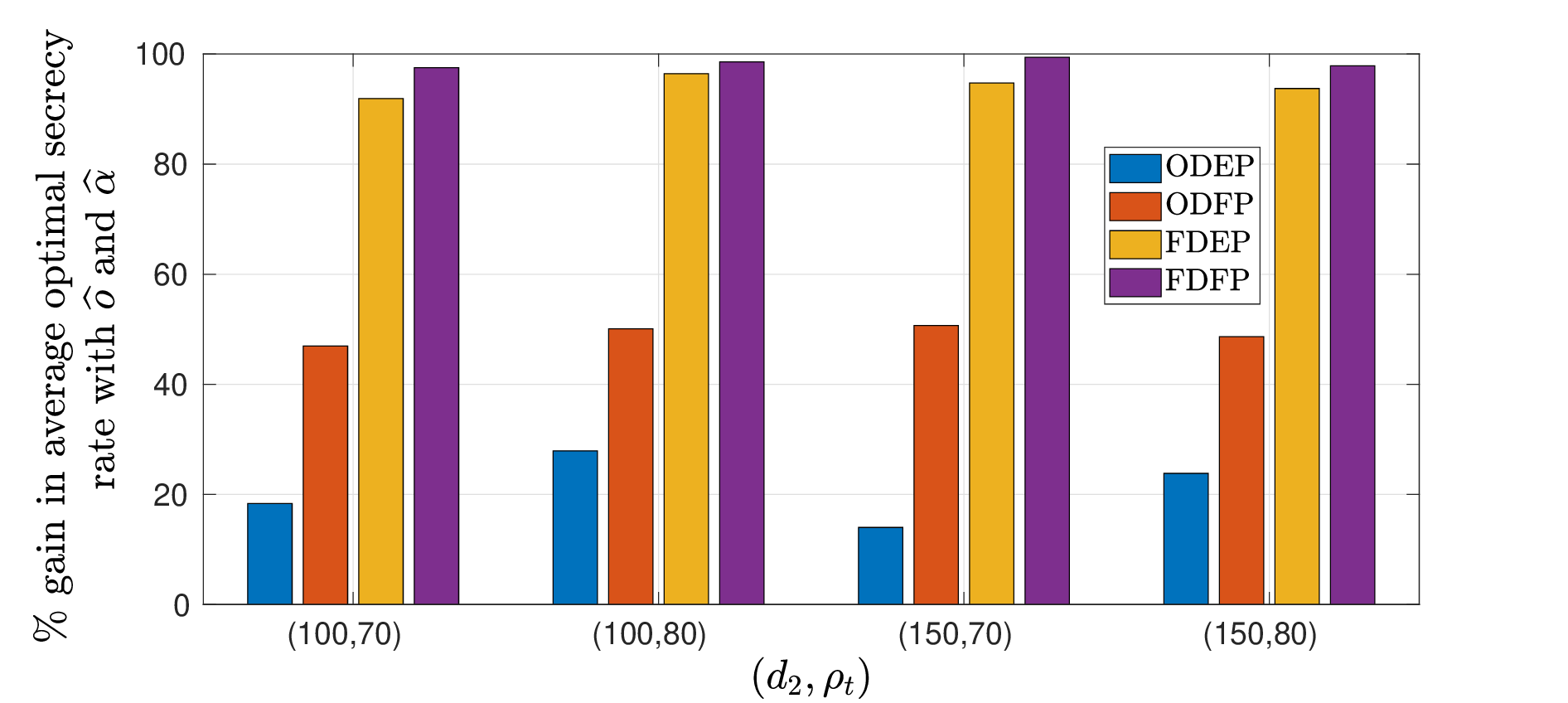}
\caption{Performance comparison of average optimal secrecy rate obtained by joint optimal solution of decoding order and power allocation, $\widehat o$ and $\widehat \alpha$, with ODEP, ODFP,  FDEP, and FDFP schemes, $\beta_{11}=\beta_{21}=0.5, \beta_{12}=0.2$. }
\label{performance_comparison}
\end{figure}

Through Fig. \ref{performance_comparison}, we demonstrate that the joint optimal solution, i.e., optimal decoding order and optimal power allocation, is capable of improving the average secrecy rate over different benchmarks. In this regard, the joint optimal solution is compared with four different benchmarks to evaluate its performance in terms of average secrecy rate, and the percentage gain is calculated. Four different benchmarks are considered: (i) ODEP: optimal decoding order and equal power allocation, (ii) ODFP: optimal decoding order and fixed power allocation, (iii) FDEP: fixed decoding order and equal power allocation, and (iv) FDFP: fixed decoding order and fixed power allocation. The optimal and fixed decoding orders are $\mathbf{D}_{2}$ and $\mathbf{D}_{4}$, respectively. 
For the equal power allocation, $\alpha=0.5$ is assumed, which is considered to examine the case in which both devices are assigned with equal power. However, in a fixed power allocation scheme, $\alpha=0.33$ is taken, which means $\alpha=0.33$ is allocated to $U_1$ and the remaining fraction $1-\alpha=0.66$ is allocated to $U_2$. Taking $\alpha=0.33$ is intended to examine the situation in which a weak device is assigned more power than a strong one, as considered in many works in the literature. Results show that the joint optimal solution provides an average percentage gain in secrecy fairness performance over benchmarks ODEP, ODFP, FDEP, and FDFP, of around $22.75\%$, $50.58\%$, $94.59\%$, and $98.16\%$, respectively. In this way, we 
 observe that the result achieved by FDFP actually differs greatly from the joint optimal solution obtained.

 \begin{figure}[!t]
\centering
\includegraphics[scale=.3]{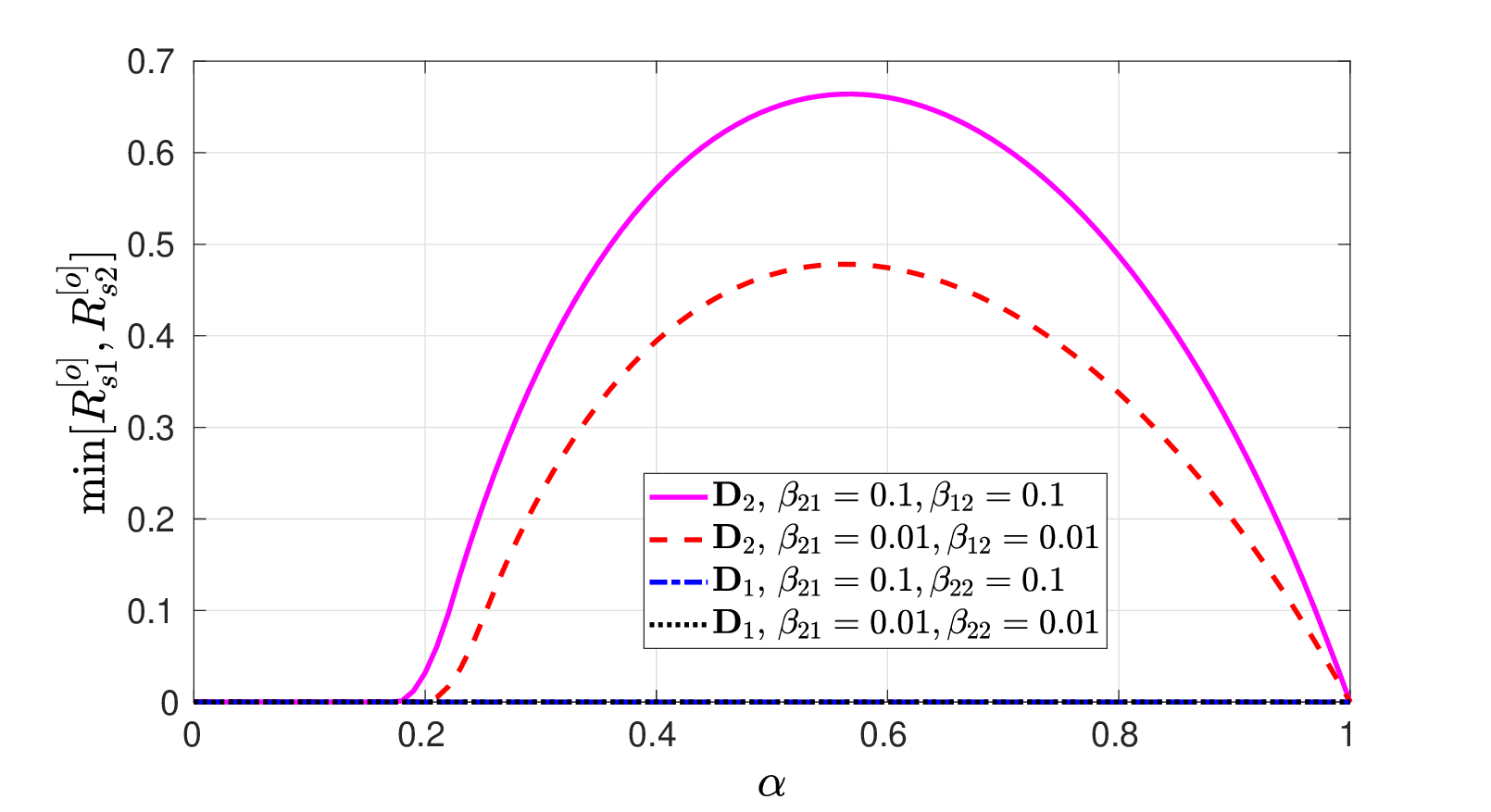}
\caption{Performance comparison of average max min secrecy  rate obtained by optimal decoding order $\mathbf{D}_{2}$ and conventional decoding order $\mathbf{D}_{1}$.}
\label{performance_comparison_2}
\end{figure}

Through Fig. \ref{performance_comparison_2}, we demonstrate a variation in $\min\left[R_{s1}^{[o]}, R_{s2}^{[o]}\right]$ with $\alpha$ for conventional decoding order and optimal decoding order. Note that the conventional and optimal secure decoding order, respectively, are $\mathbf{D}_{1}$ and $\mathbf{D}_{2}$.  Results indicate that for all $\alpha$ values, $\mathbf{D}_{2}$ ensures secrecy fairness between devices and gives a significant secrecy rate. Also, there exists a unique global-optimal solution for $\min\left[R_{s1}^{[2]}, R_{s2}^{[2]}\right]$ in terms of $\alpha$ for $\mathbf{D}_{2}$. However, in the case of conventional decoding order $\mathbf{D}_{1}$, maximizing minimum secrecy rates between devices, i.e., $\min\left[R_{s1}^{[1]}, R_{s2}^{[1]}\right]$ will always result in zero secrecy rates. The reason is that if conventional decoding order is followed, a positive secrecy rate cannot be achieved for weak devices (Refer proposition 1), resulting in  $\min\left[R_{s1}^{[2]}, R_{s2}^{[2]}\right]=0$. Thus, from the secrecy fairness viewpoint, we can conclude that $\mathbf{D}_{1}$ is not a feasible decoding order.
 
\section{Concluding Remarks}\label{section6}
We focused on obtaining a secure NOMA-enabled IIoT network with untrusted devices. We considered the RI at receivers with the linear model to observe the practical impact of imperfect SIC on the network's performance. We first obtained feasible secure decoding orders to achieve a positive secrecy rate for each device. Under each device's positive secrecy rate constraint, we jointly optimized the secure decoding order and power allocation to maximize the minimum secrecy rate between devices and provided the closed-form solution. Lastly, we presented numerical results to validate the accuracy of the theoretical analysis and provide insights into the optimal results and performance gain over benchmarks. Future work could analyze resource allocation for secrecy fairness maximization in a MIMO NOMA network with multiple devices.

\begin{appendices}
\section{Proof of Proposition 3}
In case of $\mathbf{D}_{3}=[1,2;2,1]$, both devices $U_{1}$ and $U_{2}$ first decode their own signals, perform SIC, and then decode the signal of other multiplexed device. Following Table \ref{table1}, we obtain the SINRs as
$\Gamma^{[3]}_{11} =  \frac{\alpha|h_{1}|^{2}}{(1-\alpha)|h_{1}|^{2}+\frac{1}{\rho_{t}}}$, 
$\Gamma^{[3]}_{21} = \frac{(1-\alpha)|h_{1}|^{2}}{\alpha_{1}\beta_{11}|h_{1}|^{2}+\frac{1}{\rho_{t}}}$, 
$\Gamma^{[3]}_{22} = \frac{(1-\alpha)|h_{2}|^{2}}{\alpha|h_{2}|^{2}+\frac{1}{\rho_{t}}}$, 
$\Gamma^{[3]}_{12} = \frac{\alpha|h_{2}|^{2}}{(1-\alpha)\beta_{22}|h_{2}|^{2}+\frac{1}{\rho_{t}}}$. 
Here, the condition $\Gamma^{[3]}_{11} > \Gamma^{[3]}_{12}$ for $R_{s1}^{[3]}>0$ gives a feasible condition as $\alpha > 1-\frac{|h_{1}|^{2}-|h_{2}|^{2}}{|h_{1}|^{2}|h_{2}|^{2}\rho_{t}(1-\beta_{22})}$, which shows that strong device $U_1$ can be secured from the $U_2$. However, the required condition $\Gamma^{[3]}_{22}>\Gamma^{[3]}_{21}$ for $U_{2}$ leads to $\alpha < \frac{|h_{2}|^{2}-|h_{1}|^{2}}{|h_{1}|^{2}|h_{2}|^{2}\rho_{t}(1-\beta_{11})}$, which is not a feasible condition. Thus, the data of $U_2$ is not secured against $U_1$. Hence, the decoding order $\mathbf{D}_{3}$ is infeasible.

\section{Proof of Proposition 4}
In $\mathbf{D}_{4}=[1,1;2,2]$, using Table \ref{table1}, the SINRs can be given as
$\Gamma^{[4]}_{11} = \frac{\alpha|h_{1}|^{2}}{(1-\alpha)|h_{1}|^{2}+\frac{1}{\rho_{t}}}$, 
$\Gamma^{[4]}_{21} = \frac{(1-\alpha)|h_{1}|^{2}}{\alpha\beta_{11}|h_{1}|^{2}+\frac{1}{\rho_{t}}}$, 
$\Gamma^{[4]}_{12} = \frac{\alpha|h_{2}|^{2}}{(1-\alpha)|h_{2}|^{2} +\frac{1}{\rho_{t}}}$,
$\Gamma^{[4]}_{22} = \frac{(1-\alpha)|h_{2}|^{2}}{\alpha\beta_{12}|h_{2}|^{2}+\frac{1}{\rho_{t}}}$.
To get $R^{[4]}_{s1}>0$ for $U_{1}$, we solve $\Gamma^{[4]}_{11} > \Gamma^{[4]}_{12}$, and get a feasible condition $|h_{1}|^{2}>| h_{2}|^{2}$. This shows that positive secrecy rate can be obtained for $U_{1}$ if $0<\alpha
\leq1$, since $\alpha=0$ gives $R^{[4]}_{s1}=0$. Similarly, to get  $R_{s2}^{[4]}>0$, the condition $\Gamma^{[4]}_{22}>\Gamma^{[4]}_{21}$ gives a feasible condition  $\alpha_{1}>\frac{|h_{1}|^{2}-|h_{2}|^{2}}{|h_{1}|^{2}|h_{2}|^{2}\rho_{t}(\beta_{11} - \beta_{12})}$. Thus, positive secrecy rate can be obtained for $U_2$ with a constraint on $\alpha$ as  $\frac{|h_{1}|^{2}-|h_{2}|^{2}}{|h_{1}|^{2}|h_{2}|^{2}\rho_{t}(\beta_{11} - \beta_{12})}<\alpha<1$ since $\alpha=1$ gives $R^{[4]}_{s2}=0$. Thus, to get a positive secrecy rate for both devices, the joint constraint on power allocation can be given as in \eqref{alpha_bound_D4}. 
\end{appendices}

\bibliographystyle{IEEEtran}
\bibliography{ref}

\begin{IEEEbiography}
[{\includegraphics[width=2.8cm]{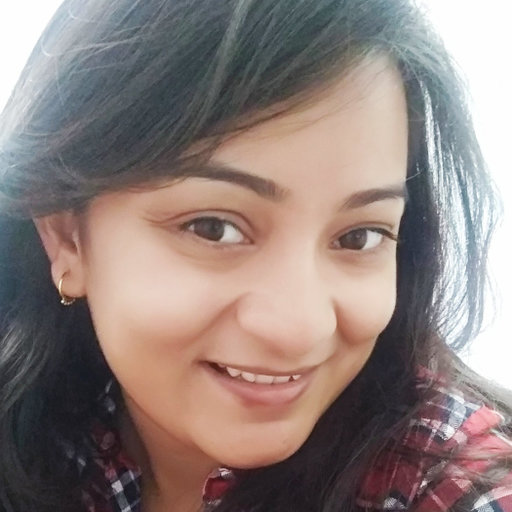}}]{Sapna Thapar} (Student Member, IEEE) received the B.Tech. degree in electronics and communication engineering from the Rajasthan Technical University, Rajasthan, India, in 2013, and the M.Tech. degree in electronics and communication engineering from the LNMIIT Jaipur, Rajasthan, India, in 2016. She is currently working toward a PhD degree in electrical engineering from the Indian Institute of Technology (IIT) Jammu, Jammu and Kashmir, India. She has also been a Visiting Researcher at the University of Tokyo, Tokyo, Japan, in 2022. She has also worked at the Lovely Professional University, Punjab, India, as an Assistant Professor, from 2016 to 2017. She is a recipient of the TCS RSP Fellowship (2019-2023). She was also selected as an Exemplary Reviewer of IEEE WIRELESS COMMUNICATIONS LETTERS in 2021. Her  research interests include non-orthogonal multiple access and physical layer security.
\end{IEEEbiography} 

\begin{IEEEbiography}
	[{\includegraphics[width=2.8cm]{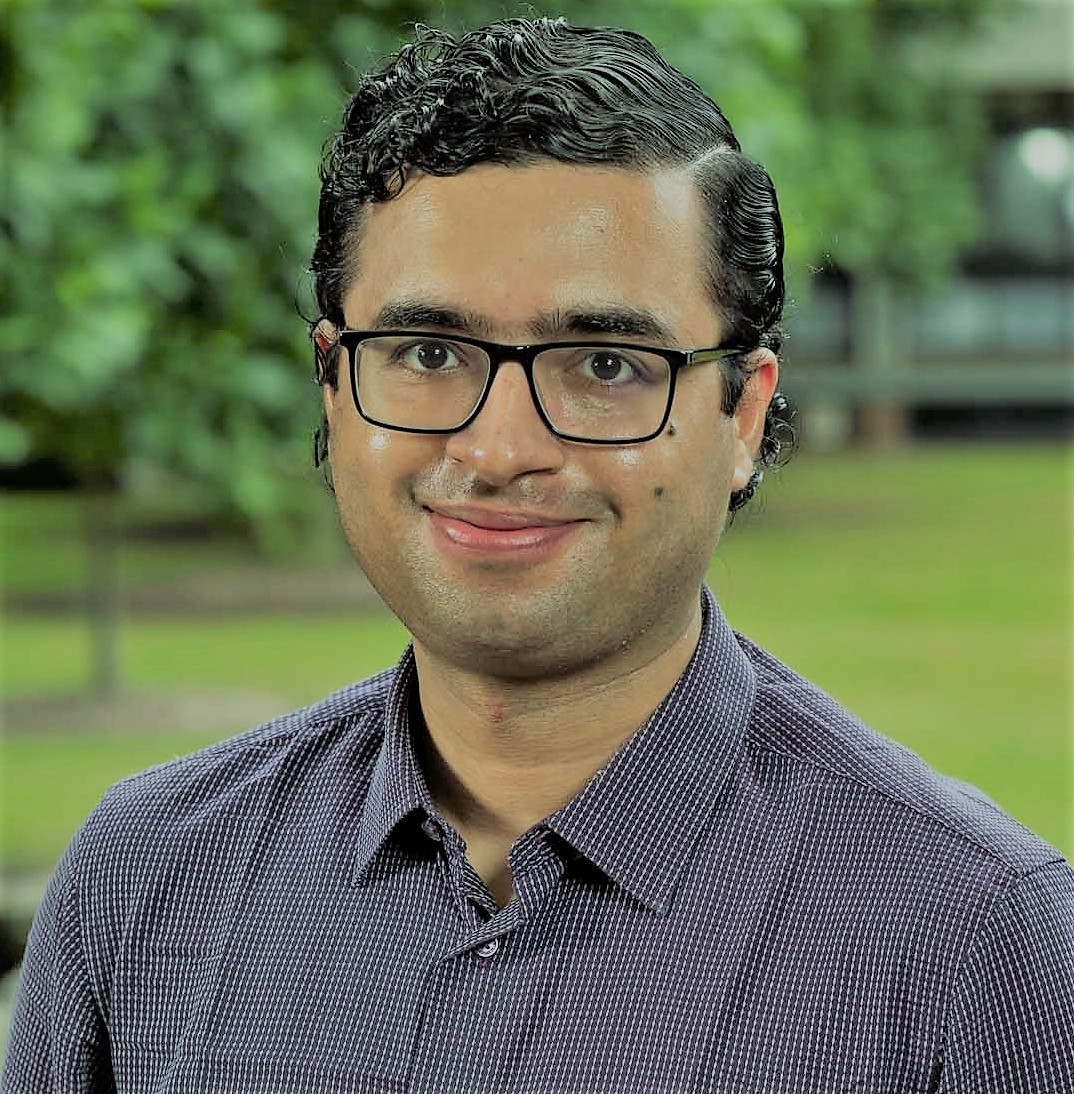}}]{Deepak Mishra} (Senior Member, IEEE) received a PhD degree in electrical engineering from the Indian Institute of Technology (IIT) Delhi in 2017. Currently, he is an Australian Research Council (ARC) Discovery Early Career Researcher Award (DECRA) fellow with the School of Electrical Engineering and Telecommunications at UNSW Sydney, where he joined in August 2019 as a Senior Research Associate. Before that, he was a Post-Doctoral Researcher at Linköping University, Sweden, from August 2017 to July 2019. He has also been a Visiting Researcher at the Northeastern University (USA), University of Rochester (USA), Huawei Technologies (France), and Southwest Jiaotong University (China). He serves as an Associate Editor of IEEE WIRELESS COMMUNICATIONS LETTERS, IEEE ACCESS, and Communication Theory track of Frontiers in Communications and Networks. His research interests include energy harvesting cooperative communication networks, MIMO, backscattering, physical layer security, as well as signal processing and energy optimization schemes for the uninterrupted operation of wireless networks.
\end{IEEEbiography} 

\begin{IEEEbiography}
	[{\includegraphics[width=2.6cm]{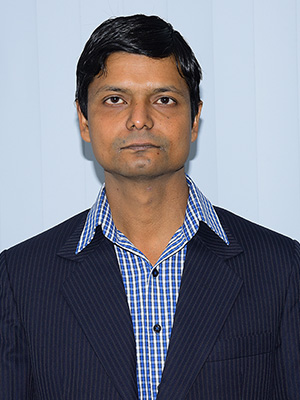}}]{Ravikant Saini} (Member, IEEE) received B.Tech. degree in electronics and communication Engineering, and M.Tech. degree in communication systems from the Indian Institute of Technology Roorkee, India, in 2001 and 2005, respectively. After completing his PhD from the Indian Institute of Technology Delhi, India, in 2016, he worked as an Assistant Professor at Shiv Nadar University, Greater Noida, India, till Dec. 2017. Since then, he has been working as an Assistant Professor in the Electrical Engineering Department at the Indian Institute of Technology (IIT) Jammu, India. 
\end{IEEEbiography} 

\end{document}